\documentclass[%
 preprint, 
 amsmath,amssymb,
 aps, physrev,
floatfix,
]{revtex4-2}
\usepackage{cleveref}
\usepackage{graphicx}
\usepackage{dcolumn}
\usepackage{bm}
\usepackage{multirow}
\usepackage{xcolor}
\colorlet{purple}{black}
\usepackage[normalem]{ulem}

\newcommand{\R}{\mathbb{R}}
\newcommand{\E}{\mathbb{E}}
\newcommand{\Var}{\mathrm{Var}}

\newcommand{\kcut}{k_{\mathrm{cut}}}

\newcommand{\what}{\widehat}

\renewcommand{\Re}{\operatorname{Re}}
\renewcommand{\Im}{\operatorname{Im}}

\begin{document}

\preprint{APS/123-QED}

\title{\textbf{\textcolor{purple}{Intrinsic instantaneous coarse-to-fine recoverability in multiscale chaos: The Lorenz--96 system}} 
}%

\author{Zhongfeng Xu}
\affiliation{
Department of Mechanical Engineering, The Hong Kong Polytechnic University, Hong Kong, China.
}%
\author{Junfeng Chen}
\email{Corresponding author: majfchen@ust.hk}
\affiliation{%
Department of Mathematics, The Hong Kong University of Science and Technology, Hong Kong, China
}%


\begin{abstract}
\textcolor{purple}{In multiscale chaotic systems, a central question for closure is how much of an unresolved fine-scale target can be inferred from a simultaneous coarse observation. For a specified observation--target pair, we quantify instantaneous recoverability under the invariant snapshot measure as the fraction of target variance explained by conditioning on the retained coarse variables. Evaluating this quantity across scales produces a scale-resolved recoverability map. Applied to the Lorenz--96 system across a range of forcings, the maps reveal a strongly nonuniform landscape: lower-wavenumber targets have little variance explained by the retained lower modes, while a finite band of higher modes exhibits partial recoverability once the retained scales reach the energetically active intermediate range. Recoverability onset is organized around a reference cutoff at which two retained modes can directly couple to the target through the quadratic nonlinearity, although its precise location depends on attractor statistics. Across the tested forcings, this qualitative onset structure persists while recoverability decreases, leaving a larger fraction of target variance unexplained in more strongly driven regimes. This definition identifies the maximal explained-variance fraction attainable by instantaneous deterministic models under squared loss; the estimated maps provide a scale-dependent diagnostic for closure modeling.}
\end{abstract}

\maketitle

\section{Introduction}
\begin{figure}[htbp]
    \centering
    \includegraphics[width=\linewidth]{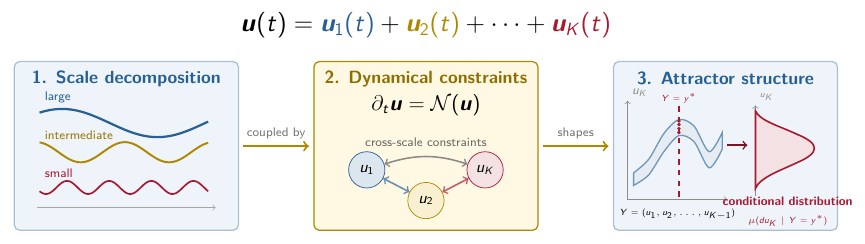}
    \caption{\textcolor{purple}{From scale decomposition to interscale structure. Coupled dynamics shape the joint statistics of scale components. Conditioning on the coarse state $Y$ yields a distribution of the fine-scale target $u_K$, whose remaining spread illustrates unresolved variability.}}
    \label{fig:interscale-structure}
\end{figure}

\textcolor{purple}{Although chaotic trajectories are sensitive to initial conditions, statistically stationary regimes can exhibit reproducible long-time statistics. In multiscale systems, this statistical organization spans interacting scales and is commonly characterized through energy spectra, correlation and structure functions, and interscale energy fluxes. Together, these descriptions motivate a broader view of interscale structure: the statistical relationships among components at different scales, shaped by their coupled dynamics (Fig.~\ref{fig:interscale-structure}). Within this broader picture, we focus on one particular aspect---instantaneous coarse-to-fine recoverability---and ask how much of a fine-scale target's variance can be explained by the coarse state at the same instant. This focus is motivated by closure modeling, where the effects of unresolved scales must be represented using resolved information.}

\textcolor{purple}{A related long-time question is addressed by classical results on determining modes and nodes, which show that finitely many observations can determine the long-time behavior of important dissipative systems.} This idea goes back to the Foias--Prodi theory for the two-dimensional Navier--Stokes equations and was later developed through determining modes, nodal values, and volume elements
\cite{foias1967comportement,foias1983number,foias1984determination,jones1992number,jones1992determining}. Related results in data assimilation and determining forms show that partial observations can synchronize or parametrize the long-time dynamics when they are used with feedback, interpolant observables, or determining ordinary differential equations \cite{azouani2014continuous,hayden2011discrete,foias2012determining,foias2014unified}.

\textcolor{purple}{Complementary statistical approaches use conditional expectations and information-theoretic measures to construct predictive closures and characterize statistical dependence. For instance, mutual information quantifies general statistical dependence, whereas transfer entropy characterizes directed, time-lagged dependence in turbulent cascades
\cite{cover2006elements,schreiber2000measuring,materassi2014information,shavit2020singular,lozano2022information}. In reduced-order dynamics, conditional expectations underpin optimal-prediction strategies like the projection-based Mori--Zwanzig formalism \cite{chorin2000optimal,chorin2002optimal,lin2023regression}. Furthermore, within turbulence modeling, stochastic and optimal estimation techniques routinely approximate conditional averages to parameterize subgrid scales \cite{adrian1989approximation,adrian1994stochastic,MoreauTeytaudBertoglio2006,berger2018numerically,Pope2010}.}

\textcolor{purple}{Here, we consider deterministic mean-square regression under the invariant snapshot measure, using only the coarse state at the same instant, without a time history or dynamical feedback. This setting is distinct from reconstruction using time-delay coordinates
\cite{takens1981detecting,casdagli1991state,abarbanel1993analysis}
and recovery through dynamical synchronization
\cite{pecora1990synchronization}.}

\textcolor{purple}{We quantify this single-time inference using the classical correlation ratio (nonlinear explained variance)
\cite{Renyi1959,HastieTibshiraniFriedman2009}. In the present study, we realize this idea with Fourier modes, linking spectral truncation directly to physical spatial resolution. Here, the retained lower modes up to a cutoff $k_{\rm cut}$ represent the macroscopic structures resolved on a coarse mesh, while a higher target mode $k$ represents the unresolved subgrid details.} For each target mode $k$ and cutoff $k_{\rm cut}$, we define $R(k\mid k_{\rm cut})$ as the variance fraction of mode $k$ explained by conditioning on the retained modes $0,\dots,k_{\rm cut}$. Evaluating this quantity over all admissible pairs produces the scale-resolved map
$$(k,k_{\rm cut})\mapsto R(k\mid k_{\rm cut}).$$  
\textcolor{purple}{For the chosen observation--target pair, recoverability gives the maximal explained-variance fraction attainable by an instantaneous deterministic regressor under squared loss. This optimum is attained by the target's conditional mean.}

We study this question in the Lorenz--96 (L96) system \cite{Lorenz1996,KarimiPaul2010}. \textcolor{purple}{The Fourier decomposition naturally aligns with the translation symmetry of L96 and makes its nonlinear mode-coupling geometry explicit, while varying the forcing parameter allows us to compare different dynamical regimes. The system therefore lets us study coarse-to-fine recovery without the geometric and modeling difficulties of more realistic fluid systems.}

The main contributions of the paper are as follows.
\begin{itemize}
    \item We formulate a conditional-variance-based recoverability map $R(k\mid k_{\rm cut})$ under the invariant snapshot measure.
    \item We apply this map to Fourier-mode observations in the Lorenz--96 system. For $F=8,16,32,64$, we estimate $\widehat R(k\mid k_{\rm cut})$ for all target modes and all allowed lower-mode cutoffs. The results show a finite-band structure that changes with forcing. The highest modes are the most recoverable in all four regimes, while stronger forcing lowers their recoverability and leaves more conditional variance unexplained.
    \item We support the maps with internal checks, including saturation with respect to data size and network width, the population monotonicity check, and onset curves for the lowest cutoff at which each target mode becomes recoverable.
    \item We introduce an RFF residual check to test whether recoverable structure remains after the fitted regressor. \textcolor{purple}{On representative mode pairs, the check finds little residual predictability within the tested probe classes, providing evidence for estimator adequacy}.
\end{itemize}
\section{Lorenz--96 as a coarse-to-fine inverse problem}
\label{sec:l96}

\subsection{Dynamics}

We consider the standard $N$-variable Lorenz--96 system, 
a widely used model for spatially extended nonlinear dynamics 
\cite{kerin2022lorenz,van2018travelling,li2025direct}:
\begin{equation}
\dot{x}_j
=
(x_{j+1}-x_{j-2})x_{j-1} - x_j + F,
\qquad
j=1,\dots,N,
\label{eq:l96}
\end{equation}
with periodic boundary conditions:
\begin{equation}
x_{j+N}=x_j.
\end{equation}
\textcolor{purple}{Here, $x_j$ represents a macroscopic scalar state variable at the $j$-th discrete spatial location. Its evolution is governed by a constant external forcing $F$ that injects energy, a linear damping term $-x_j$ representing internal dissipation, and a nonlinear advection term $(x_{j+1}-x_{j-2})x_{j-1}$ that conserves total energy while driving its transfer across spatial scales.}

In the main numerical study, we take $N=40$ and $F\in\{8,16,32,64\}$. 
These choices place the system in regimes of sustained spatiotemporal chaos, where the attractor exhibits an extensive character \cite{KarimiPaul2010}. Crucially, the nonlinear term in \eqref{eq:l96} is strictly quadratic. Consequently, in the Fourier representation introduced below, it couples modes exclusively through triadic interactions. 
This explicit geometric constraint makes the Lorenz--96 system an ideal testbed for studying how nonlinear mode interactions shape cross-scale statistical relations under the invariant measure.

\subsection{Fourier representation}
\textcolor{purple}{We use a Fourier decomposition as it matches the translation symmetry of the Lorenz--96 system, in which Fourier modes provide a natural global ordering by spatial wavenumber. A lower-wavenumber cutoff therefore defines a simple nested hierarchy of coarse observations. More importantly, in Fourier coordinates the quadratic nonlinearity obeys the exact triadic selection rule \(p+q\equiv k\pmod N\), which provides the dynamics-based reference scale used below (see Section~\ref{subsec:triad}). The recoverability definition itself is not restricted to Fourier observations.}

We define the discrete Fourier coefficients of the state as
\begin{equation}
\hat{x}_k
=
\frac{1}{\sqrt{N}}
\sum_{j=1}^{N}
x_j \exp\!\left(-2\pi i \frac{(j-1)k}{N}\right),
\qquad
k=0,\dots,N-1.
\label{eq:dft}
\end{equation}
Since the state is real-valued, the coefficients satisfy the conjugate symmetry $\hat{x}_{N-k}=\overline{\hat{x}_k}$. Thus, for $N=40$, it is enough to keep the modes $k=0,1,\dots,20$. The zero mode $k=0$ is real. The Nyquist mode $k=N/2=20$ is also real. The modes $k=1,\dots,19$ are complex. These modes contain $1+2\cdot 19+1=40=N$ real degrees of freedom, which matches the dimension of the original state.

For target modes $k=1,\dots,19$, we use the real and imaginary parts as a real-valued target vector:
\begin{equation}
U_k
=
\bigl(\Re \hat{x}_k,\; \Im \hat{x}_k\bigr)\in\R^2.
\label{eq:Uk}
\end{equation}
For the Nyquist mode $k=20$, the imaginary part is always zero, so we define
\begin{equation}
U_{20}=\Re \hat{x}_{20}\in\R.
\label{eq:U20}
\end{equation}
Thus the dimension of the target is
\begin{equation}
d_k=
\begin{cases}
2, & 1\le k\le 19,\\
1, & k=20,
\end{cases}
\end{equation}
and $U_k\in\R^{d_k}$ for all target modes.

\subsection{Observation operator}

For a target mode $k$, we let the regressor use only lower modes. Thus the cutoff satisfies
\begin{equation}
\kcut\in\{0,1,\dots,k-1\},
\end{equation}
where $\kcut$ is the largest retained mode index. The observation contains the modes
\begin{equation}
0,1,\dots,\kcut.
\end{equation}
The zero mode is real, so we keep only its real part. For $\kcut\ge 1$, we define the observation vector as:
\begin{equation}
Y_{\kcut}
=
\Bigl(
\Re\hat{x}_0,\,
\Re\hat{x}_1,\Im\hat{x}_1,\,
\dots,\,
\Re\hat{x}_{\kcut},\Im\hat{x}_{\kcut}
\Bigr)\in\R^{2\kcut+1}.
\label{eq:obs}
\end{equation}
For the base case $\kcut=0$, the observation is the scalar $Y_0=\Re\hat{x}_0\in\R$. With this notation established, the core inverse problem is exactly posed:
\begin{quote}
\textit{Given the retained lower modes $Y_{\kcut}$, how much of the target mode $U_k$ at the same instant can be recovered?}
\end{quote}
\section{Population definition of instantaneous recoverability}
\label{sec:framework}

We start with the inverse problem before choosing any learning model. Let $\mu$ be a probability measure on the finite-dimensional Hilbert space $\mathcal{H}$. At the population level, $\mu$ is the invariant snapshot measure of the dynamical system. In computations, we approximate $\mu$ by the empirical measure from a long trajectory after transients. Let $U$ be an $\mathcal{H}$-valued random state with law $\mu$ and

\begin{equation}
Y = D(U)
\end{equation}
be the coarse observation, where $D$ is a known observation map, and let
\begin{equation}
U_B = \Pi_B U
\end{equation}
be the target unresolved component, where $\Pi_B$ projects onto a chosen band, shell, or mode. All expectations and conditional expectations below are taken under the law of $(U_B,Y)$ given by $\mu$. We assume $U_B\in L^2(\Omega;\mathcal{H})$ and $\Var(U_B)=\E\|U_B-\E U_B\|_2^2>0$.

\subsection{The optimal deterministic regressor and recoverability}
\label{subsec:definition}

\textcolor{purple}{We consider square-integrable deterministic regressors
$g(Y)$ of $U_B$ from the simultaneous observation $Y$, assuming
$\E\|U_B\|_2^2<\infty$.
The optimal regressor $m_B^\star$ minimizes the population mean-square error:}
{\color{purple}
\begin{equation}
m_B^\star
=
\operatorname*{arg\,min}_{g:\,\E\|g(Y)\|_2^2<\infty}
\E\|U_B-g(Y)\|_2^2.
\label{eq:prediction_problem}
\end{equation}
A standard result in regression analysis shows that this minimizer is
unique up to almost-sure equality and is given by the conditional
mean~\cite[Sec.~2.4]{HastieTibshiraniFriedman2009}
\cite[Sec.~1.5.5]{bishop2006pattern}:
\begin{equation}
m_B^\star(Y)=\E[U_B\mid Y].
\label{eq:condmean}
\end{equation}
}
It is also the $L^2$ projection of $U_B$ onto the closed space of
square-integrable functions of $Y$.
\textcolor{purple}{This optimal regressor yields the decomposition}
{\color{purple}
\begin{equation}
U_B
=
\underbrace{m_B^\star(Y)}_{
    \substack{\text{optimal deterministic}\\\text{recovery}}
}
+
\underbrace{\epsilon_B}_{
    \substack{\text{remaining conditional}\\\text{variability}}
},\qquad\text{with}\quad
\E[\epsilon_B\mid Y]=0.
\label{eq:optimal_recovery_decomposition}
\end{equation}
}
\textcolor{purple}{The corresponding variance decomposition is}
\begin{equation}
\E\|U_B-\E U_B\|_2^2
=
\underbrace{\E\|m_B^\star(Y)-\E U_B\|_2^2}_{\text{explained}}
+
\underbrace{\E\|U_B-m_B^\star(Y)\|_2^2}_{\text{residual}}.
\label{eq:variance_decomposition}
\end{equation}
\textcolor{purple}{For a target with nonzero variance, we define the \emph{instantaneous recoverability}
of $U_B$ from $Y$ as the explained-variance fraction}
\begin{equation}
R_B
=
\frac{\E\|m_B^\star(Y)-\E U_B\|_2^2}
     {\E\|U_B-\E U_B\|_2^2}.
\label{eq:RB_explained_variance}
\end{equation}
Equations~\eqref{eq:variance_decomposition}--\eqref{eq:RB_explained_variance}
imply $0\leq R_B\leq 1$.
Values near zero indicate that $Y$ explains little of the variance
of $U_B$, whereas values near one indicate near-deterministic recovery.

\textcolor{purple}{All expectations are taken under the joint law of $(U_B,Y)$ induced by $(\mu,D,\Pi_B)$, so $R_B$ is a population property intrinsic to this selected triple. It quantifies deterministic mean-square recoverability, without describing the full conditional law of $U_B$ given $Y$.}

\subsection{Monotonicity and regression estimation}
\label{subsec:monotonicity}

We will use two basic facts from the projection view. First, recoverability increases when more information is observed. Since the conditional expectation is the $L^2$ projection onto the space of functions of the observation, conditioning on a finer observation cannot increase the residual variance:
\begin{equation}
\sigma(Y_1)\subseteq\sigma(Y_2)
\quad\Longrightarrow\quad
R_B(Y_1)\le R_B(Y_2).
\label{eq:monotonicity}
\end{equation}
This monotonicity holds exactly at the population level. In Section~\ref{sec:estimation}, we use it to check the quality of the estimator.

\textcolor{purple}{Second, for any candidate regressor $g$, define the
population score attained by $g$ as
\begin{equation}
R_B[g]
=
1-
\frac{\E\|U_B-g(Y)\|_2^2}
{\E\|U_B-\E U_B\|_2^2}.
\label{eq:predictor_score}
\end{equation}
By the projection property of the conditional mean,
\begin{equation}
R_B-R_B[g]
=
\frac{\E\|g(Y)-m_B^\star(Y)\|_2^2}
{\E\|U_B-\E U_B\|_2^2}
\ge 0.
\label{eq:predictor_gap}
\end{equation}
Thus $R_B[g]\le R_B$, with equality only when
$g(Y)=m_B^\star(Y)$ almost surely. The regression architecture determines how closely an estimation approaches the population optimum. The numerical implementation and its adequacy checks are described in Section~\ref{sec:estimation}.}

\subsection{Modewise recoverability}
\label{subsec:modewise}

In the Lorenz--96 setting introduced in Section~\ref{sec:l96}, we take the target $B$ to be a single Fourier mode $k$. We take the observation to be the lower modes kept up to a cutoff $\kcut<k$. We write
\begin{equation}
R(k\mid \kcut)
\label{eq:Rkkcut}
\end{equation}
for the recoverability of mode $k$ from the retained modes up to $\kcut$. This gives a modewise version of Equation~\eqref{eq:RB_explained_variance}, indexed by the target mode and the cutoff. Computing these entries over all admissible pairs gives the recoverability map studied below.
\section{Results}
\label{sec:results}
\textcolor{purple}{Throughout this section, $\widehat R$ denotes the recoverability score computed on held-out data using the fitted regressor. The regression procedure and estimator-adequacy checks are described in Section~\ref{sec:estimation}.}
\subsection{Recoverability map}
\begin{figure}[t]
    \centering
    \includegraphics[width=0.92\textwidth]{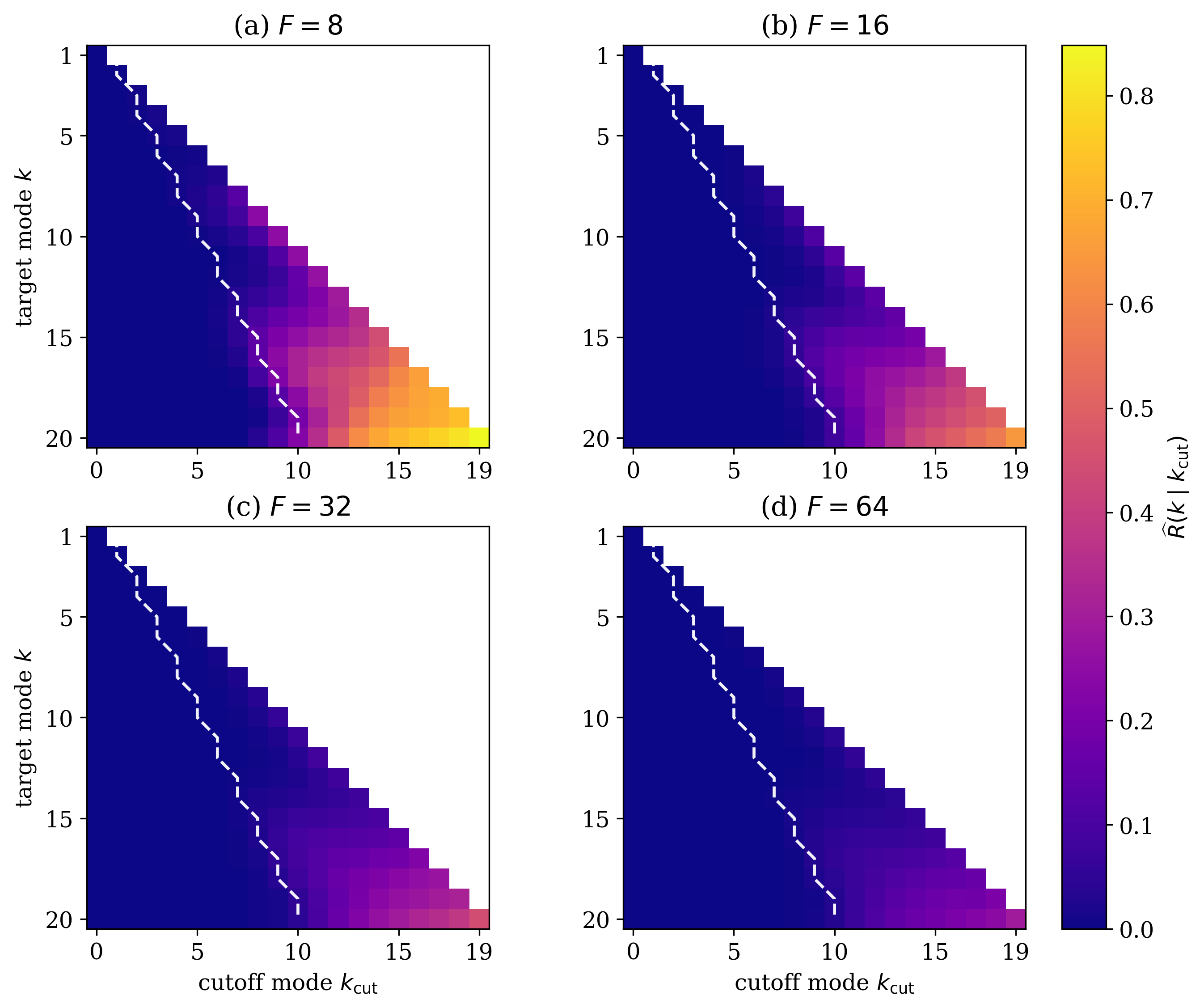}
    \caption{Empirical recoverability map $\what{R}(k\mid \kcut)$ for
    the Lorenz--96 system with $N=40$ and $F=8,16,32,64$. The vertical
    axis is the target mode index $k$, and the horizontal axis is the
    largest retained mode index $\kcut$. The region $\kcut\ge k$ is
    masked because the target mode is already included in the
    observation. A common color scale is used across all four forcings.
    The dashed curve marks the quadratic access scale
    $\kcut=\lceil k/2\rceil$.}
    \label{fig:heatmap}
\end{figure}

Figure~\ref{fig:heatmap} shows the empirical recoverability score $\what{R}(k\mid \kcut)$ for all admissible pairs $(k,\kcut)$ and all four forcing values. The map is lower triangular in the $(k,\kcut)$ plane. We do not estimate the region $\kcut\ge k$, because the target mode is already included in the observation there and the recoverability is equal to one.

The maps show a clear structure, and two features appear for all four forcing values. First, recoverability varies strongly across the allowed pairs $(k,\kcut)$. Low target modes remain weakly constrained by still coarser observations, while clear recoverability appears in a finite band of high target modes. For a fixed target mode, recoverability usually increases as more lower modes are retained. This agrees with the population monotonicity property indicated in Equation~\eqref{eq:monotonicity}, up to the small empirical fluctuations reported later in Table~\ref{tab:monotonicity}. Second, recoverability decreases as the forcing increases. The near-diagonal high-mode value $\widehat R(20\mid19)$ drops from $0.8485$ at $F=8$ to $0.2964$ at $F=64$. The same decrease is also visible across the broader high-mode band. Thus, increasing $F$ keeps the same overall geometric pattern, but reduces the fraction of the high-mode variance removed by conditioning on the retained modes.

This trend can be explained by the dynamics. Larger forcing makes the system more chaotic, with stronger nonlinear transfer and larger conditional variation across Fourier modes. The retained modes still take part in the quadratic interactions that shape the high-mode statistics, but they constrain the high modes less strongly, so conditioning removes a smaller fraction of their variance.

This decrease is also visible across the whole map. To measure it, we summarize each map by the maximum recoverability
\[
R_{\max}(F)=\max_{k,\kcut}\what{R}_F(k\mid\kcut),
\]
the map mean \(M(F)\), averaged over all 210 admissible pairs, and the active area fraction
\[
A_{0.2}(F)
=
\frac{1}{210}
\#\{(k,\kcut):\what{R}_F(k\mid\kcut)>0.2\}.
\]
Here \(A_{0.2}(F)\) is the fraction of pairs for which conditioning removes more than 20\% of the target-mode variance. As shown in Table~\ref{tab:landscape_summary}, all three quantities decrease as the forcing increases. The maximum recoverability coincides with the near-diagonal value $\widehat R(20\mid 19)$ reported above. The map mean decreases from 0.1368 to 0.0277. The active area fraction shrinks even more, from 0.2667 to 0.0238. Thus stronger forcing reduces both the peak recoverability and the number of mode--cutoff pairs where deterministic closure from the retained modes has a clear effect.

\begin{table}[!htbp]

\centering
\caption{Scalar summaries of the empirical recoverability maps:
the maximum recoverability \(R_{\max}\), the landscape mean \(M\), and
the active area fraction \(A_{0.2}\) of admissible pairs achieving more than a 20\% variance reduction.}
\label{tab:landscape_summary}
\begin{ruledtabular}
\begin{tabular}{cccc}
\(F\) & \(R_{\max}\) & $M$ & \(A_{0.2}\) \\
\colrule
8  & 0.8485 & 0.1368 & 0.2667 \\
16 & 0.6447 & 0.0759 & 0.1524 \\
32 & 0.4448 & 0.0454 & 0.0857 \\
64 & 0.2964 & 0.0277 & 0.0238 \\
\end{tabular}
\end{ruledtabular}
\end{table}

The finite high-mode band therefore identifies where lower modes carry usable instantaneous information about high modes. Its weakening with $F$ indicates that stronger forcing leaves a larger unresolved component after conditioning on the chosen retained variables. In closure terms, this means that the same retained variables become less informative as the forcing increases.

\begin{figure}[t]
    \centering
    \includegraphics[width=0.5\textwidth]{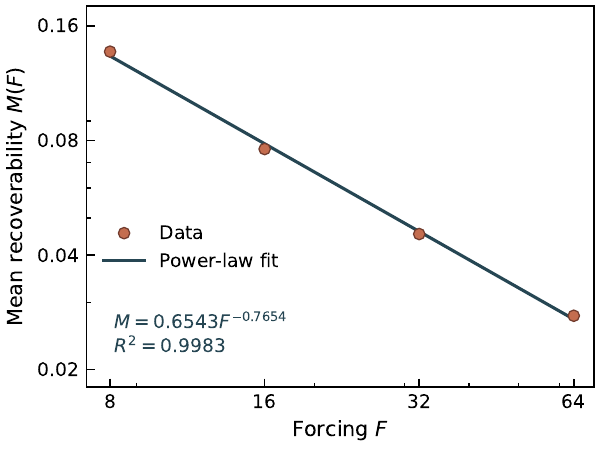}
    \caption{Landscape-averaged recoverability $M(F)$ versus forcing $F$ on a log–log scale.}
    \label{fig:power_law}
\end{figure}

\textcolor{purple}{Beyond the pointwise decrease of individual entries, the landscape-averaged recoverability $M(F)$ itself follows an approximate power-law decay with forcing. As shown in Figure~\ref{fig:power_law}, a log–log regression over the four available values yields
\begin{equation*}
M(F)\;\propto\;F^{-\alpha},\qquad\alpha\approx0.7654
\end{equation*}
($R^2=0.9983$). While the limited number of points precludes a definitive exponent, the clean scaling quantifies how the overall amplitude of instantaneous recoverability diminishes under stronger driving.}

\subsection{Direct triad access as a reference scale for recoverability growth}
\label{subsec:triad}

The heat maps in \Cref{fig:heatmap} show how recoverability changes as the cutoff $\kcut$ increases. To understand where this growth starts along the cutoff axis, we first use the quadratic Fourier coupling of the Lorenz--96 system to define a simple reference scale.

Using the Fourier convention in Equation~\ref{eq:dft} and setting $\theta_m=2\pi m/N$, a direct calculation gives the following equation for the Fourier modes, with all mode indices taken modulo $N$:
\begin{equation}
\frac{d}{dt}\hat{x}_k
=
\frac{1}{\sqrt{N}}
\sum_{\substack{p+q\equiv k\,(\mathrm{mod}\,N)}}
C_{p,q}\,\hat{x}_p\hat{x}_q
-\hat{x}_k+\sqrt{N}F\delta_{k0},
\label{eq:l96_fourier}
\end{equation}
where
\begin{equation}
C_{p,q}
=
\exp\!\bigl(i(\theta_p-\theta_q)\bigr)
-
\exp\!\bigl(-i(2\theta_p+\theta_q)\bigr).
\label{eq:triad_coefficient}
\end{equation}
\textcolor{purple}{The nonlinear term is quadratic because each contribution is proportional to a product $\hat{x}_p\hat{x}_q$. It therefore couples the Fourier modes in triads: only pairs $(p,q)$ satisfying $p+q\equiv k\pmod N$ can contribute directly to the evolution of mode $k$.}

For a cutoff $\kcut$, the observation contains the positive modes $1,\dots,\kcut$. Since the state is real-valued, it also determines the negative modes $-\kcut,\dots,-1$ by conjugate symmetry, together with the zero mode. Thus a direct quadratic interaction from retained modes to target mode $k$ can only use pairs
\begin{equation}
p,q\in \mathcal{K}_{\kcut}:=\{-\kcut,\dots,\kcut\}.
\end{equation}
For the target modes used here, $1\le k\le 20=N/2$. If $2\kcut<k$, then no pair in $\mathcal{K}_{\kcut}$ can satisfy $p+q\equiv k\pmod N$. Indeed, the ordinary sum $p+q$ lies in $[-2\kcut,2\kcut]$, so it cannot equal $k$. The aliased case $p+q=k-N$ is also outside this range. Therefore, the first cutoff at which two retained modes can directly reach mode $k$ through a quadratic triad is
\begin{equation}
\kcut=\left\lceil\frac{k}{2}\right\rceil.
\label{eq:half_scale}
\end{equation}
For even $k$, this is realized by the pair $(k/2,k/2)$. For odd $k$, it is realized by $\bigl((k-1)/2,(k+1)/2\bigr)$. For $k=1$, this formal cutoff is outside the allowed range $\kcut<k$, so it is used only as part of the reference curve.

The scale $\lceil k/2\rceil$ is a property of the instantaneous vector field, and serves as a reference scale rather than a sharp threshold, since recoverability is a property of the invariant measure.

To summarize this growth, we define the required cutoff at a given threshold. For a threshold $\varepsilon>0$, let
\begin{equation}
\what{k}_{\mathrm{cut}}^{\mathrm{req}}(k;\varepsilon)
=
\min\{\kcut\in\{0,\dots,k-1\}:\what{R}(k\mid \kcut)>\varepsilon\},
\label{eq:kcutreq}
\end{equation}
when this set is nonempty. If no allowed cutoff gives recoverability above the threshold, we omit that target mode. We use $\varepsilon=10^{-2}$ to remove near-zero values while still keeping weak recoverability. We also use $\varepsilon=5\times 10^{-2}$ as a stricter threshold for clearer recoverability.

\begin{figure}[t]
    \centering
    \includegraphics[width=0.95\textwidth]{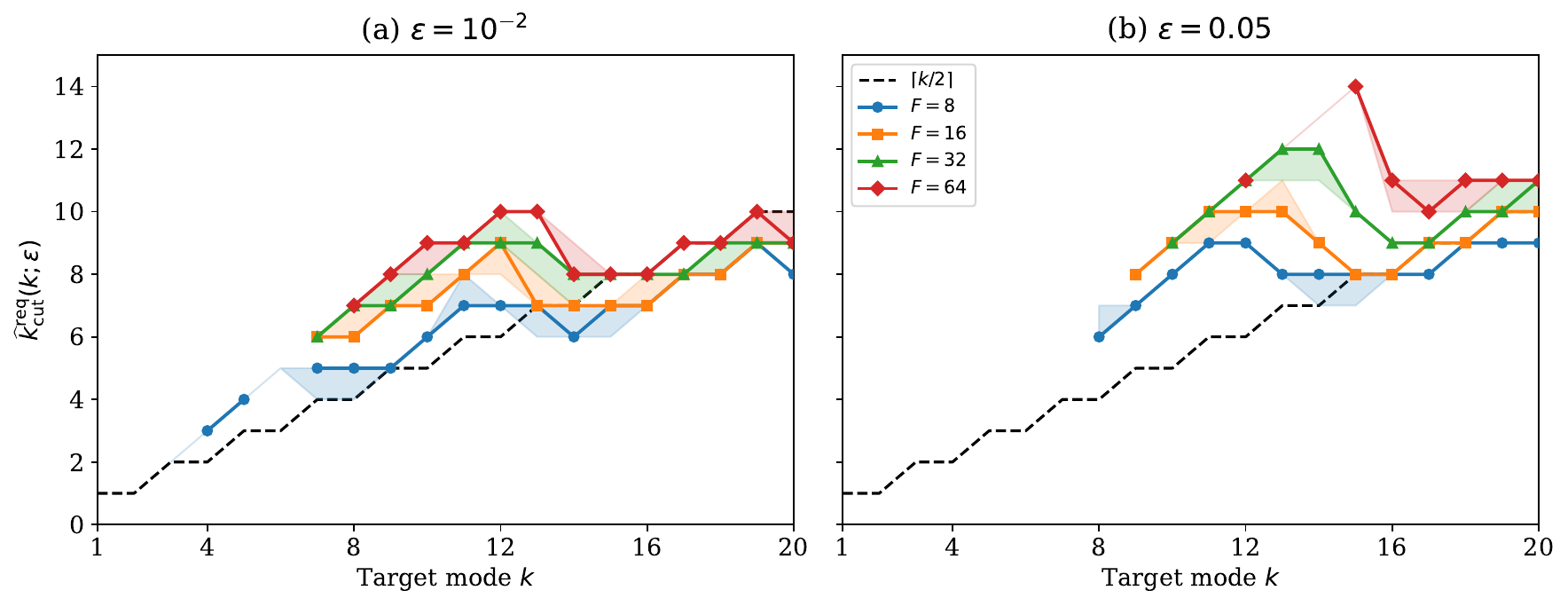}
    \caption{Thresholded required cutoff $\what{k}_{\mathrm{cut}}^{\mathrm{req}}(k;\varepsilon)$ for each target mode $k$ and forcings $F=8,16,32,64$. $(a)$ and $(b)$ correspond to $\varepsilon=10^{-2}$ and $\varepsilon=5\times 10^{-2}$. Missing points indicate that no recoverability score of admissible lower-mode cutoff exceeds the indicated threshold. Shaded bands show bootstrap confidence intervals. The dashed curve marks the direct retained--retained quadratic triad-access scale $\kcut=\lceil k/2\rceil$.}
    \label{fig:required_cutoff}
\end{figure}

Figure~\ref{fig:required_cutoff} shows that the empirical required cutoffs are organized around the direct-access scale, but they do not follow it exactly. For $\varepsilon=10^{-2}$, the curves stay close to $\lceil k/2\rceil$ over much of the range shown. For low and intermediate target modes, the required cutoff is often larger than the direct-access scale. This means that the first available quadratic route is not always enough to give clear recoverability, for example when the modes in these triads have little energy or weak statistical coupling. For higher target modes, some crossings occur below $\lceil k/2\rceil$. This suggests that the retained modes can already contain weak indirect information about the target before a direct retained--retained triad is available.

When we raise the threshold to $\varepsilon=5\times 10^{-2}$, many weak crossings disappear and the remaining curves move upward. Low and intermediate target modes often have no crossing at this stricter threshold. The remaining points tend to occur at cutoffs that include the energetic intermediate modes. The dependence on forcing is also clear. Stronger forcing usually requires a larger cutoff to reach the same threshold, and fewer modes pass the stricter test.

Thus the required cutoff reflects two effects. The scale $\lceil k/2\rceil$ gives the first cutoff at which retained modes can directly reach mode $k$ through a quadratic interaction. The empirical crossing also depends on the energy in the modes, the strength of statistical coupling on the attractor, the forcing value, and the chosen threshold. Direct triad access gives the main geometric pattern, while the actual onset of recoverability is set by the statistics of the invariant measure.
\section{Numerical methods and estimator diagnostics}
\label{sec:estimation}
\textcolor{purple}{We now describe the data construction, regression estimator, and diagnostics underlying Figures~\ref{fig:heatmap}--\ref{fig:required_cutoff}.}
\subsection{Data generation and sampling}
We generate the data by directly integrating Equation~\eqref{eq:l96} from the standard near-equilibrium initial condition
\begin{equation}
x_1(0)=F+0.01,\quad \text{and}\quad x_j(0)=F\quad \text{for}\quad j=2,\dots,N.
\end{equation}

For each value of $F$, we integrate the system with a fixed-step fourth-order Runge--Kutta method with a time step of $\Delta t=0.01$.
To remove the effect of the initial transient, we discard the first $2,001$ time steps as burn-in, yielding a transient time of $T_{\mathrm{burn}}=20$.
After that, we integrate the system for another $5\times10^8$ time steps. We keep one snapshot every $100$ integration steps, so that
\begin{equation}\label{eq:100subsample}
\Delta t_{\rm samp}=100\Delta t=1.
\end{equation}
Thus the retained trajectory covers about $5.0\times10^6$ time units after burn-in.

The recoverability quantity in Section~\ref{sec:framework} is an expectation under the invariant snapshot measure. In the computations, we estimate this expectation by averaging over the retained snapshots. For this estimate to be reliable, the retained snapshots should sample one stationary distribution and should not be too strongly correlated in time. We check these two points in turn below.

\begin{figure}[!t]
    \centering
    \includegraphics[width=0.45\textwidth]{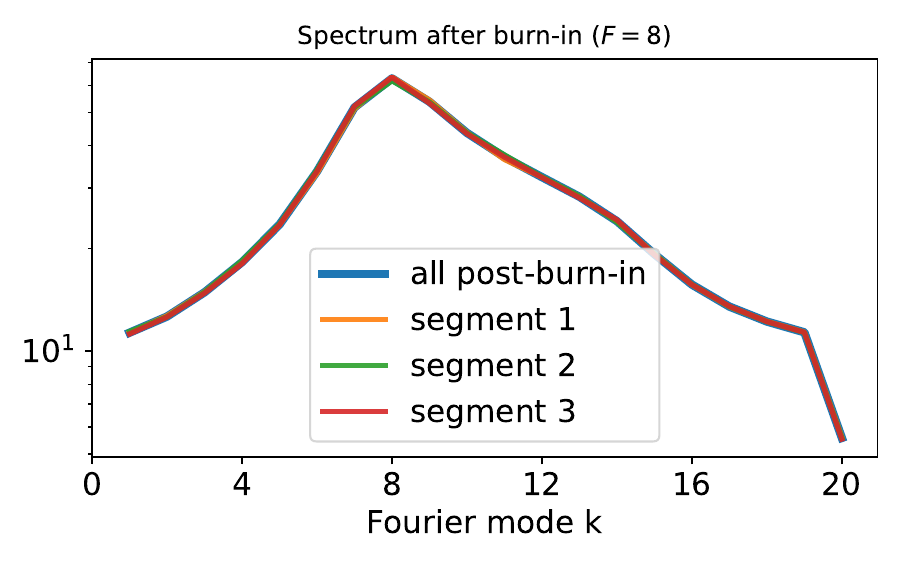}\hfill
    \includegraphics[width=0.45\textwidth]{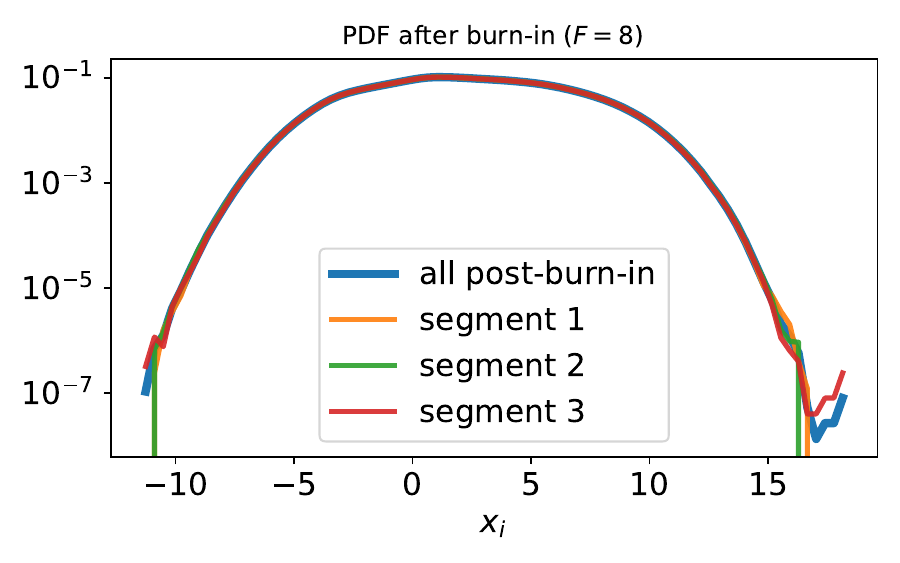}\\
    \includegraphics[width=0.45\textwidth]{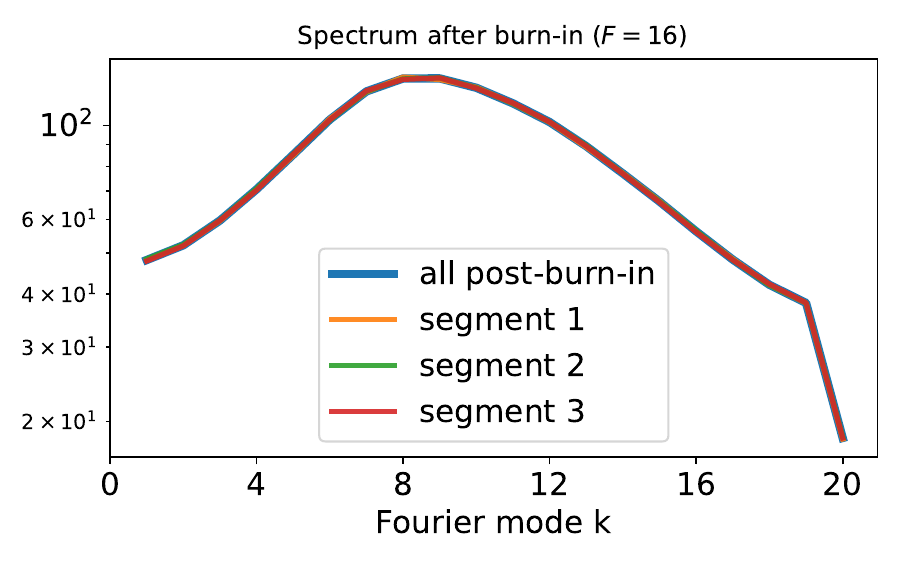}\hfill
    \includegraphics[width=0.45\textwidth]{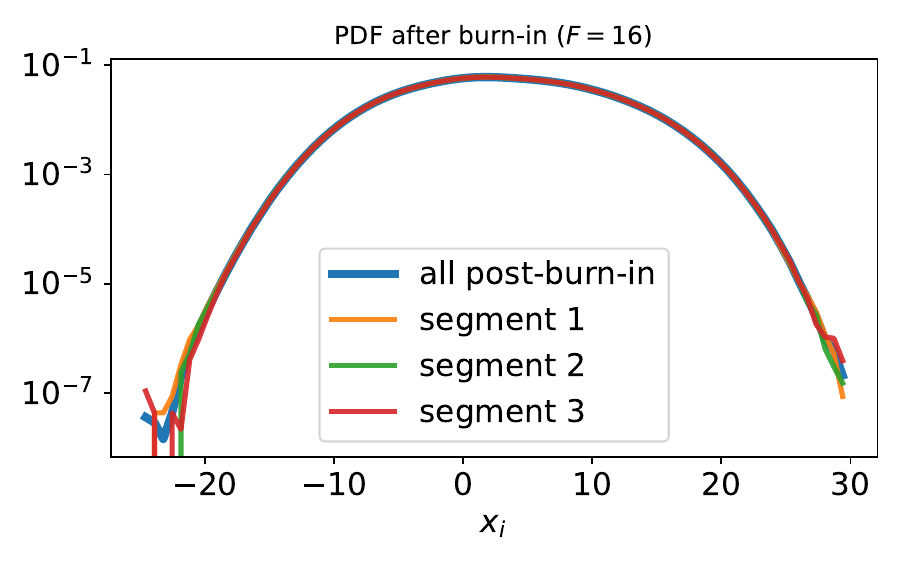}\\
    \includegraphics[width=0.45\textwidth]{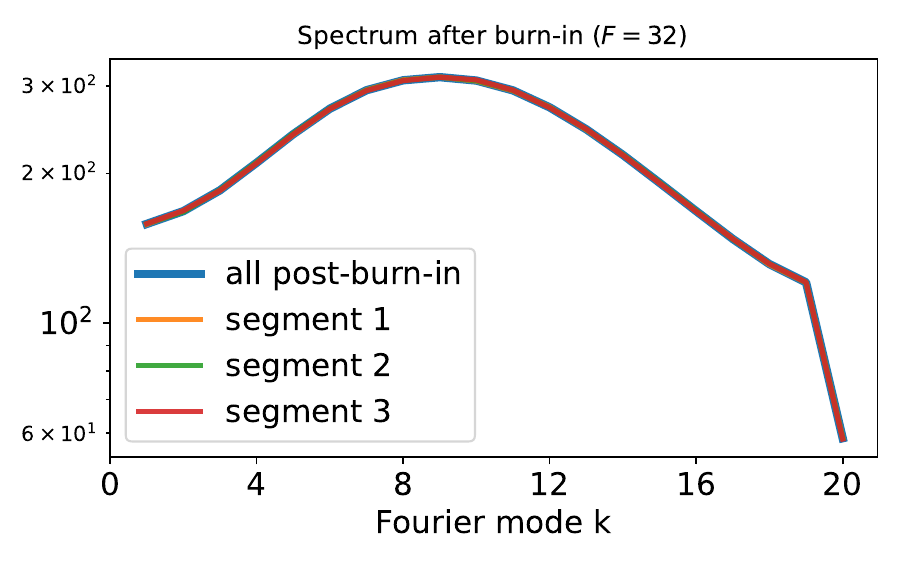}\hfill
    \includegraphics[width=0.45\textwidth]{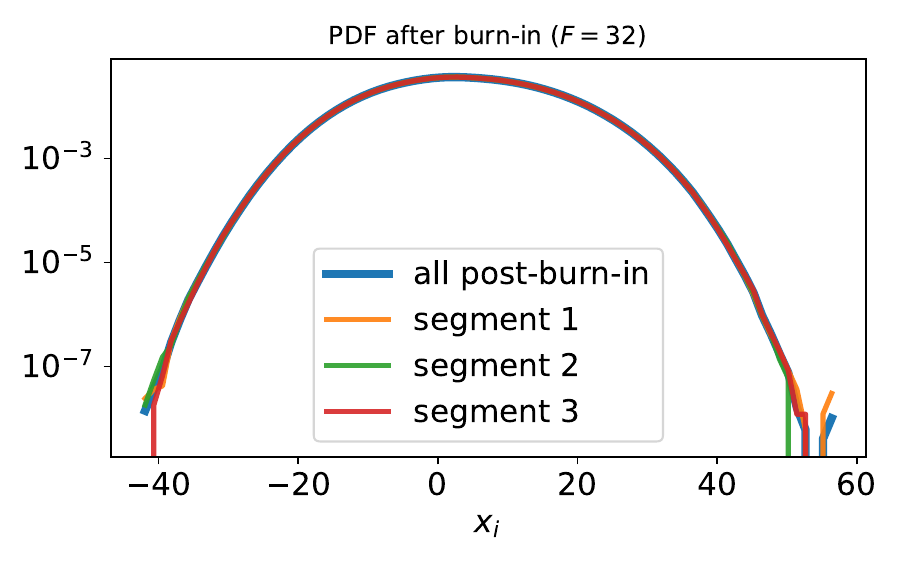}\\
    \includegraphics[width=0.45\textwidth]{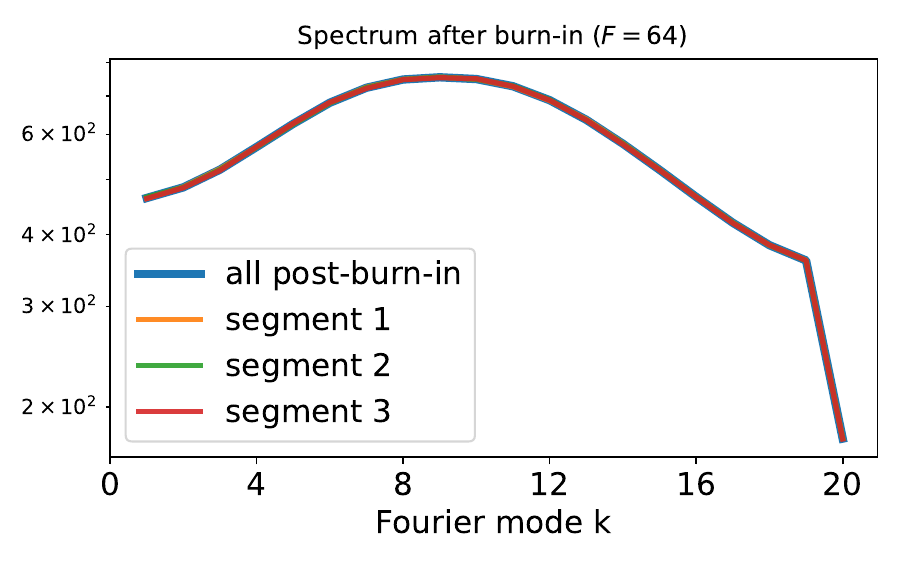}\hfill
    \includegraphics[width=0.45\textwidth]{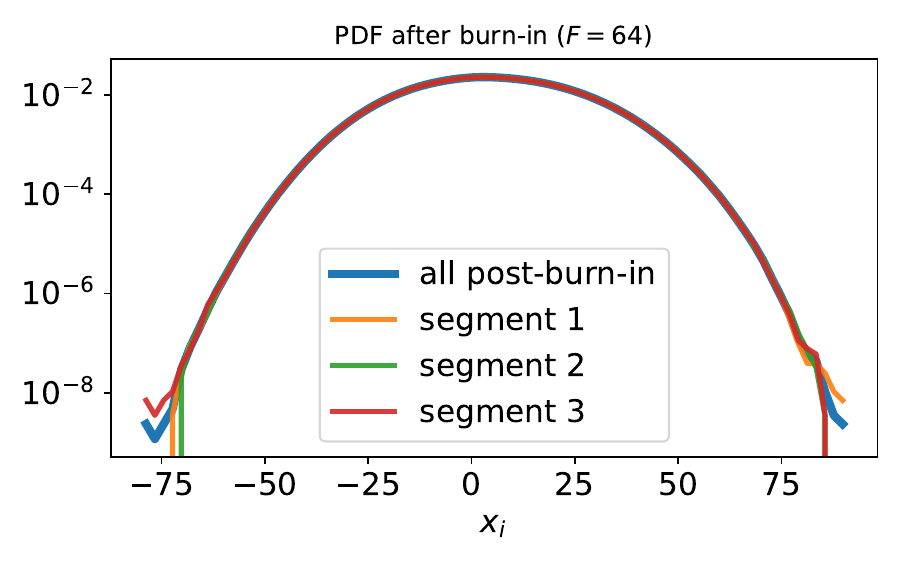}
    \caption{Stationarity check: each post-burn-in trajectory is split into three equal-length consecutive segments, and the spatial Fourier power spectrum (left) and pooled one-point density (right) are compared across segments. The close agreement of the segment-wise curves indicates statistical stationarity; the spectra share the same intermediate-wavenumber peak and decay, while the densities agree except in the far tails, where sampling noise is largest.}
    \label{fig:data}
\end{figure}

\paragraph{Stationarity.}
We split each trajectory after burn-in into three consecutive segments of equal length and compare two statistics across them: the spatial Fourier power spectrum and the pooled one-point density of the state variables; see Figure~\ref{fig:data}. For each value of $F$, the curves from the three segments are almost the same. This shows that, after burn-in, each trajectory samples a stable statistical regime with no clear slow drift. The plots also show that the four forcing values lead to different regimes. As $F$ increases, the spectral power becomes larger and the one-point density spreads over a wider range of state values. At the same time, the main spectral peak remains in the intermediate range $k=8$--$10$.

\paragraph{Temporal subsampling.}
The raw outputs are separated by only $\Delta t_{\rm raw}=0.01$, so neighboring raw states are strongly correlated in time. If we used all raw outputs, the sample size would look much larger, but many samples would carry nearly the same information. We therefore subsample the trajectory in time. The goal is to keep snapshots that are closer to independent samples from the stationary distribution. We choose the sampling interval using autocorrelation. For a complex Fourier coefficient $\hat x_k(t)$, we compute the normalized autocorrelation
\begin{equation}
\rho_k(\tau)
=
\frac{
\left\langle
\bigl(\hat x_k(t)-\langle \hat x_k\rangle\bigr)
\overline{\bigl(\hat x_k(t+\tau)-\langle \hat x_k\rangle\bigr)}
\right\rangle
}{
\left\langle
\left|\hat x_k(t)-\langle \hat x_k\rangle\right|^2
\right\rangle
}.
\label{eq:autocorr}
\end{equation}
Here, the brackets mean averaging over a trajectory after burn-in. Since $\hat x_k$ is complex-valued, we use $\operatorname{Re}\rho_k(\tau)$ to measure how fast the mode loses correlation in time. For several representative Fourier modes, we report two time scales: the first time $\tau_{1/e}$ at which $\operatorname{Re}\rho_k(\tau)$ falls below $e^{-1}$, and the integrated autocorrelation time
\begin{equation}
\tau_{\rm int}
=
\Delta t_{\rm raw}
\left(
1+2\sum_{m=1}^{m_0-1}
\operatorname{Re}\rho_k(m\Delta t_{\rm raw})
\right)
\label{eq:tauint}
\end{equation}
in Table~\ref{tab:decorrelation}. Here, $m_0$ is the first index such that $\operatorname{Re}\rho_k(m\Delta t_{\rm raw})\leq 0$. The same autocorrelation curves are shown in Figure~\ref{fig:decorrelation} of Appendix~\ref{app:decorrelation}. Among all tested forcings and modes, the largest $1/e$ decorrelation time is $\tau_{1/e}=0.40$, and the largest integrated autocorrelation time is $\tau_{\rm int}=0.681$. Both occur at the weakest forcing $F=8$. Thus the retained interval $\Delta t_{\rm samp}=1$ chosen in Equation~\ref{eq:100subsample}  is larger than the main short-time correlation scales. For stronger forcing, the modes decorrelate faster, so the same sampling interval is even safer.

\begin{table}[ht!]
\centering
\caption{Temporal decorrelation diagnostics for representative Fourier modes $k=0,5,10,15,20$. For each forcing, the table reports the largest $1/e$ decorrelation time, the mode attaining it, and the largest integrated autocorrelation time. The raw output interval is $\Delta t_{\rm raw}=0.01$; the retained interval after subsampling is $\Delta t_{\rm samp}=1$.}
\label{tab:decorrelation}
\begin{ruledtabular}
\begin{tabular}{ccccc}
$F$ & $\max_k \tau_{1/e}$ & slowest mode for $\tau_{1/e}$ & $\max_k \tau_{\rm int}$ & $\Delta t_{\rm samp}/\max_k \tau_{\rm int}$ \\
\colrule
$8$  & $0.40$ & $k=10$ & $0.681$ & $1.47$ \\
$16$ & $0.22$ & $k=10$ & $0.398$ & $2.51$ \\
$32$ & $0.14$ & $k=0$  & $0.233$ & $4.28$ \\
$64$ & $0.09$ & $k=0$  & $0.148$ & $6.74$ \\
\end{tabular}
\end{ruledtabular}
\end{table}
\subsection{From conditional-mean approximation to empirical recoverability}

Suppose we have snapshot pairs
\begin{equation}
\left\{Y_{\kcut}^{(i)}, U_k^{(i)}\right\},
\qquad
i=1,\dots,N_s.
\end{equation}
For each pair $(k,\kcut)$, we fit a map
\begin{equation}
g_{\theta,k}^{(\kcut)}:\R^{2\kcut+1}\to\R^{d_k}
\end{equation}
by minimizing the squared loss on the training set:
\begin{equation}
\mathcal{L}_{k,\kcut}(\theta)
=
\frac{1}{|I_{\rm train}|}
\sum_{i\in I_{\rm train}}
\left\|
g_{\theta,k}^{(\kcut)}(Y_{\kcut}^{(i)})
-
U_k^{(i)}
\right\|_2^2.
\label{eq:loss}
\end{equation}
By Section~\ref{sec:framework}, the population minimizer of this squared loss is the conditional mean $m_k^\star(Y_{\kcut})=\E[U_k\mid Y_{\kcut}]$. In our computations, we take $g_{\theta,k}^{(\kcut)}$ to be a one-hidden-layer neural network. Standard universal-approximation results show that neural networks with nonpolynomial activations can approximate broad classes of continuous functions on compact sets \cite{cybenko1989approximation,hornik1989multilayer,leshno1993multilayer}. With enough data and enough network capacity, the trained network is expected to approximate
\begin{equation}
g_{\theta,k}^{(\kcut)}(Y_{\kcut}) \;\approx\; m_k^\star(Y_{\kcut})=\E[U_k\mid Y_{\kcut}].
\end{equation}
\textcolor{purple}{It is worth noting that the trained networks are only used to estimate the pairwise recoverability score, and never used as closures or inserted back into the filtered dynamical systems. These scores diagnose how much of the unresolved-scale variance can be explained using the resolved variables under the invariant measure.}
\subsection{Empirical recoverability score}
In computation, we replace the expectations in $R(k\mid \kcut)$ with averages over a finite set of snapshots. This is reasonable for the data used here. After burn-in, the Lorenz--96 trajectory is in a statistically stable regime. Under the standard ergodic view, long-time averages along the trajectory approximate expectations under the invariant snapshot distribution.

Let
\begin{equation}
\bar{U}_k
=
\frac{1}{|I_{\rm test}|}
\sum_{i\in I_{\rm test}} U_k^{(i)}
\end{equation}
be the mean of the target mode on the test set. We replace the conditional mean $\E(U_k\mid Y_{\kcut})$ by the trained model $g_{\theta,k}^{(\kcut)}(Y_{\kcut})$. This gives the empirical recoverability score
\begin{equation}
\what{R}(k\mid \kcut)
=
1
-
\frac{
\sum_{i\in I_{\rm test}}
\left\|
U_k^{(i)} - g_{\theta,k}^{(\kcut)}(Y_{\kcut}^{(i)})
\right\|_2^2
}{
\sum_{i\in I_{\rm test}}
\left\|
U_k^{(i)} - \bar{U}_k
\right\|_2^2
}.
\label{eq:Rhat}
\end{equation}
\textcolor{purple}{Equation~\eqref{eq:predictor_gap} shows that this score is a lower bound on the intrinsic recoverability.} Small negative estimates, when they occur, are treated as finite-sample or optimization fluctuations and are not assigned physical meaning.

\subsection{Estimation setup}
After burn-in removal and temporal subsampling, we split the post-transient snapshots of each trajectory into three disjoint blocks: a fitting set, an internal validation set, and a held-out test set. We denote by $K=N_{\rm fit}+N_{\rm val}$ the total size of the fitting and validation sets. In all reported runs, the validation set contains $20\%$ of $K$, so that $N_{\rm fit}=0.8K$ and $N_{\rm val}=0.2K$. The fitting set is used to train the network weights. The validation set is used to choose the reported checkpoint. The held-out test set is not used for training or checkpoint selection.

For each target mode $k$ and allowed cutoff $\kcut\in\{0,\dots,k-1\}$, we train a one-hidden-layer fully connected network to predict $U_k$ from $Y_{\kcut}$ by minimizing the squared loss. The hidden width is $1000$, and the activation function is GELU~\cite{hendrycks2016gaussian}. Each network is first trained with Adam~\cite{kingma2014adam} for $100{,}000$ gradient steps. We use a batch size of $16{,}384$ and an initial learning rate of $10^{-2}$, which is reduced to $10^{-4}$ with a cosine annealing scheduler~\cite{loshchilov2017sgdr}. Starting from the best Adam checkpoint, we further refine the network with chunked full-batch L-BFGS. All recoverability values in the heat maps and summary tables are computed on the held-out test block. All networks are implemented in Python with PyTorch~\texttt{2.5.1+cu121} and trained on a single NVIDIA GeForce RTX 4090 GPU.

\subsection{Diagnostics for estimator adequacy}
\label{subsec:diagnostics}

We use three diagnostics to assess estimator error: monotonicity in the retained cutoff, residual orthogonality, and saturation with respect to data size and network width.

\paragraph{Monotonicity.}
The first check is monotonicity. By Equation~\eqref{eq:monotonicity}, the population map must be nondecreasing in $\kcut$ for each fixed target mode $k$. Thus, if the empirical map decreases after one more mode is added to the observation, we treat the decrease as an error from finite data, optimization, or approximation. For each adjacent pair, we define
\begin{equation}
\Delta^{\rm mono}_{k,\kcut}
=
\left[
\what{R}(k\mid \kcut)
-
\what{R}(k\mid \kcut+1)
\right]_+,
\qquad
0\le \kcut\le k-2,
\label{eq:mono_delta}
\end{equation}
where $[a]_+=\max\{a,0\}$. We summarize these decreases by
\begin{equation}
\resizebox{0.90\linewidth}{!}{%
$\displaystyle
\begin{aligned}
V_{\max}^{\rm mono}
&=\max_{k,\kcut}\Delta_{k,\kcut}^{\rm mono},
&
V_{\rm avg}^{\rm mono}
&=\frac{1}{N_{\rm adj}}
  \sum_{k,\kcut}\Delta_{k,\kcut}^{\rm mono},
&
f_{\rm viol}^{\rm mono}
&=\frac{1}{N_{\rm adj}}
  \#\{(k,\kcut):\Delta_{k,\kcut}^{\rm mono}>0\}.
\end{aligned}
$%
}
\label{eq:mono_summary}
\end{equation}
where the sums are over all allowed adjacent cutoff pairs. We report these values only as estimator-quality checks and do not force monotonicity by isotonic regression or other post-processing.

\paragraph{Random Fourier feature (RFF) residual check.}
A more stringent check uses the $L^2$ projection property of the conditional mean. If $g(Y_{\kcut})=\E[U_k\mid Y_{\kcut}]$, then the residual $e_k=U_k-g(Y_{\kcut})$ is orthogonal to every square-integrable function of the retained observation. That is, for each component $\ell$ and every probe $\psi\in L^2(P_{Y_{\kcut}})$,
\begin{equation}
\E\!\left[e_{k,\ell}\,\psi(Y_{\kcut})\right]=0.
\label{eq:orthogonality_condition_empirical}
\end{equation}
If the residual is still correlated with some function of the retained modes, then the regressor has missed some recoverable information and has not reached the conditional mean. We cannot check Equation~\eqref{eq:orthogonality_condition_empirical} for all possible functions using finite data. We therefore test it on many random Fourier feature~\cite{rahimi2007random} (RFF) probes, following the residual-orthogonality idea in the conditional-mean barrier framework \cite{chen2026diagnosing}. We use RFF probes instead of polynomial dictionaries for two reasons. First, a fixed-degree polynomial dictionary can only detect dependence up to that degree, while the residual may contain more general structure. Second, polynomial dictionaries become too large in the present setting. For example, for the pair $(\kcut,k)=(19,20)$, the input dimension is $d=2\kcut+1=39$. A total-degree polynomial dictionary with degree $p=2,3,4$ would contain $820$, $11{,}480$, and $123{,}410$ monomials, respectively. RFF probes avoid this growth: the number of probes $D_m$ can be chosen independently of the input dimension.

On the held-out diagnostic set, we standardize each coordinate of the retained observation and write the standardized observation as $\widetilde Y_{\kcut}$. For a lengthscale $\gamma>0$, we draw $w_j\sim N(0,\gamma^{-2}I)$ and $b_j\sim\mathrm{Unif}[0,2\pi]$, and define the RFF probes
\begin{equation}
\psi_j(\widetilde Y_{\kcut})
=
\sqrt{2}\,\cos\!\left(w_j^\top \widetilde Y_{\kcut}+b_j\right),
\qquad
j=1,\dots,D_m.
\label{eq:rff_probe}
\end{equation}
These probes are bounded and square-integrable, so they are valid test functions in Equation~\eqref{eq:orthogonality_condition_empirical}. As $D_m$ increases, their span becomes dense in $L^2(P_{Y_{\kcut}})$~\cite{rahimi2007random}. Thus, if the residual is orthogonal to all such probes, then it has no remaining $L^2$ component that can be written as a function of the retained observation.

We check orthogonality to these probes using two quantities. The primary one measures how much of the residual the probes can explain together. Since the probes are correlated and each probe may explain only a small part of the remaining signal, we do not test the probes one by one. Instead, we fit a cross-validated ridge regression from the RFF probes to the residual. We then compute the out-of-fold explained variance
\begin{equation}
e_{\rm joint}^{\rm RFF}
=
1-
\frac{
\sum_{i}\bigl\|e_k^{(i)}-\widehat e_{k,{\rm oof}}^{(i)}\bigr\|_2^2
}{
\sum_{i}\|e_k^{(i)}\|_2^2
},
\label{eq:rff_joint_effect}
\end{equation}
where $\widehat e_{k,{\rm oof}}^{(i)}$ is the out-of-fold ridge prediction of the residual. If $e_{\rm joint}^{\rm RFF}\le\epsilon$, with threshold $\epsilon=0.01$, we treat the residual as having no clear recoverable structure left.

The auxiliary quantity counts how many probes are individually correlated with the residual beyond sampling noise. For each probe, we compute a studentized residual--feature statistic, and use a Gaussian multiplier bootstrap at level $\alpha=0.05$ to set a threshold for the largest statistic over the whole probe family. A probe is called \emph{significant} if its statistic is above this threshold, and $|S|$ denotes the number of significant probes. Using the largest statistic controls the family-wise error while avoiding a separate test for each probe. The count $|S|$ is only a supporting check. The main criterion is the joint quantity in Equation~\eqref{eq:rff_joint_effect}. If $e_{\rm joint}^{\rm RFF}\le\epsilon$, we accept the residual even if one isolated probe is flagged. The full definitions of both quantities, the bootstrap procedure, and the probe settings are given in Appendix~\ref{app:rff}.

As a simple bias check, we also report separately the mean residual—the case $\psi\equiv1$ of Equation~\eqref{eq:orthogonality_condition_empirical}—in Appendix~\ref{app:rff}.

\paragraph{Capacity and data-size saturation.}
We use two more checks to see whether the reported scores are stable with respect to network size and data size. For network size, we increase the hidden width of the one-hidden-layer network until the prediction error and recoverability score stop changing much. We then choose a final width in this stable range, so the estimates are not limited by a network that is too small. For data size, we recompute the map using larger and larger parts of the training trajectory, while keeping the test set fixed. If the recoverability scores become stable for large training sets, this suggests that the map is not mainly caused by finite-sample error. We carry out both checks for a representative high-mode pair and report the results below.

\subsection{Estimator-quality checks}
We use the diagnostics in Section~\ref{subsec:diagnostics} to check whether the reported maps mainly describe the inverse problem, instead of limits of the estimator.

\begin{figure}[t]
    \centering
    \includegraphics[width=0.49\textwidth]{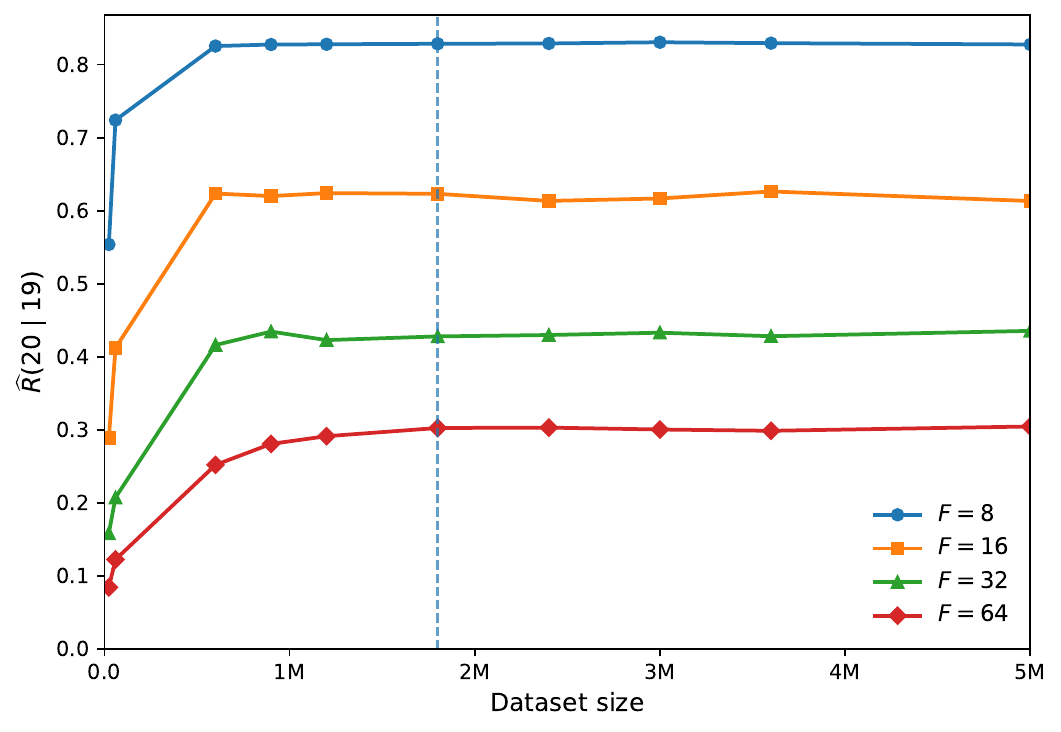}
    \hfill
    \includegraphics[width=0.49\textwidth]{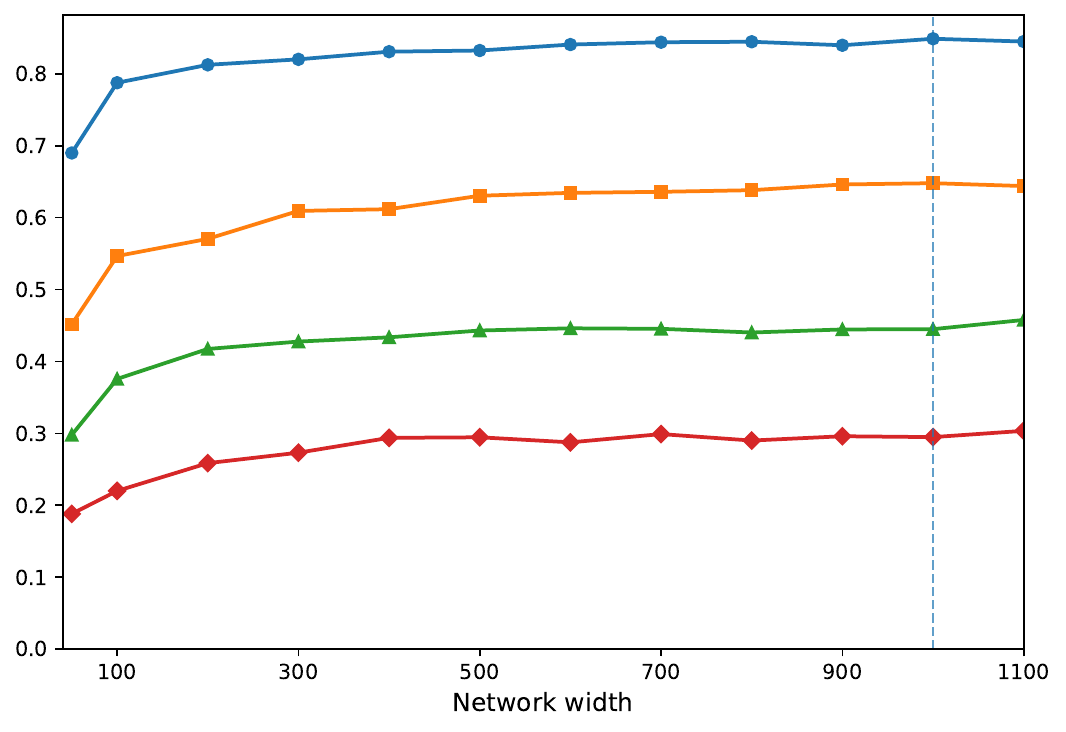}
    \caption{Estimator-saturation diagnostics for the representative
    high-mode pair $(\kcut,k)=(19,20)$. Left: dataset-size dependence of
    $\widehat{R}(20\mid 19)$ for $F=8,16,32,64$; the dashed line marks the dataset size selected for the final maps, $1.8$M. Right:
    hidden-width dependence for the same pair. The curves show practical
    stabilization with increasing dataset size and network width.}
    \label{fig:saturation_diagnostics}
\end{figure}

First, we vary the dataset size for the representative near-diagonal high-mode pair $(\kcut,k)=(19,20)$ for all four forcing values. As shown in Figure~\ref{fig:saturation_diagnostics}a, the scores increase quickly when the dataset is small and then become stable. The ordering across the four forcing values is also stable for all tested dataset sizes. This supports the dataset size used for the final maps, shown by the dashed line at $1.8$M.

Second, we vary the hidden width while keeping the dataset size fixed. As shown in Figure~\ref{fig:saturation_diagnostics}b, the scores increase when the width is small and then become stable for larger widths. The ordering across the four forcing values remains stable in this large-width range, and we do not see a clear improvement beyond widths of order $10^3$. Some small non-monotone changes remain, especially for the intermediate forcing cases. These changes are likely due to optimization and finite-sample variation, not to a change in the main ordering. We therefore use hidden width $1000$ for the final maps.

Third, we apply the RFF residual diagnostic from Section~\ref{subsec:diagnostics} to several representative near-diagonal pairs, $(\kcut,k)\in\{(19,20),(14,15),(10,11),(5,6)\}$, for all four forcing values. For each pair, the held-out residual is
\begin{equation}
e_k=U_k-g_{\theta,k}^{(\kcut)}(Y_{\kcut}).
\end{equation}
The check uses $D_m=24{,}000$ RFF probes. We use three lengthscales: the median-distance scale, one half of it, and one quarter of it. The significance level is $\alpha=0.05$, and the tolerance is $\epsilon=0.01$. The full settings are given in Appendix~\ref{app:rff}. The results are summarized in Table~\ref{tab:rff_residual_diagnostics}.

\begin{table}[ht!]
\centering
\caption{RFF residual diagnostics for representative near-diagonal pairs. For each pair, we report the empirical recoverability score $\widehat R(k \mid k_{\rm cut})$, the components of the mean residual $\widehat\mu_{e,\Re}$ and $\widehat\mu_{e,\Im}$ defined in Equation~\eqref{eq:mean_residual} (the imaginary component is absent for the real Nyquist mode $k=20$), and, over the tested RFF settings (Appendix~\ref{app:rff}), the largest number of bootstrap-significant probes $\max|S|$ and the largest cross-validated joint explained variance $\max e^{\rm RFF}_{\rm joint}$. The recoverability score and the RFF diagnostic are evaluated on the held-out test block. The joint explained variance is clipped to $[0,1]$, so zeros denote non-positive raw out-of-fold values.}
\label{tab:rff_residual_diagnostics}
\setlength{\tabcolsep}{14pt}
\renewcommand{\arraystretch}{1.15}
\begin{ruledtabular}
\begin{tabular}{ccccccc}
$F$ & pair & $\widehat R(k\mid k_{\rm cut})$ & $\widehat\mu_{e,\Re}$ & $\widehat\mu_{e,\Im}$ & $\max |S|$ & $\max e_{\rm joint}^{\rm RFF}$ \\
\colrule
\multirow{4}{*}{8} & $20\mid19$ & 0.8485 & 1.85e-05 & -- & 1 & 6.15e-05 \\
 & $15\mid14$ & 0.4427 & 2.25e-04 & 5.77e-05 & 1 & 0 \\
 & $11\mid10$ & 0.2525 & -1.26e-04 & 2.26e-04 & 0 & 0 \\
 & $6\mid5$ & 0.0088 & 1.56e-04 & -5.32e-04 & 0 & 7.24e-04 \\
\colrule
\multirow{4}{*}{16} & $20\mid19$ & 0.6447 & 2.56e-04 & -- & 0 & 0 \\
 & $15\mid14$ & 0.1986 & -5.71e-04 & 2.39e-04 & 0 & 0 \\
 & $11\mid10$ & 0.1299 & -2.89e-04 & -1.18e-04 & 0 & 0 \\
 & $6\mid5$ & 0.0046 & -1.12e-04 & 2.19e-04 & 0 & 0 \\
\colrule
\multirow{4}{*}{32} & $20\mid19$ & 0.4448 & -3.42e-05 & -- & 0 & 0 \\
 & $15\mid14$ & 0.1024 & -2.40e-04 & 5.01e-05 & 0 & 1.67e-04 \\
 & $11\mid10$ & 0.0728 & -9.01e-05 & -5.35e-05 & 0 & 0 \\
 & $6\mid5$ & 0.0041 & 1.04e-04 & 3.80e-04 & 0 & 0 \\
 \colrule
\multirow{4}{*}{64} & $20\mid19$ & 0.2964 & -3.13e-05 & -- & 0 & 0 \\
 & $15\mid14$ & 0.0583 & -5.45e-05 & -5.91e-05 & 0 & 0 \\
 & $11\mid10$ & 0.0448 & -2.61e-05 & -9.57e-06 & 1 & 1.17e-04 \\
 & $6\mid5$ & 0.0036 & -9.72e-05 & -2.19e-04 & 0 & 0 \\
\end{tabular}
\end{ruledtabular}
\end{table}

Across all tested  pairs and RFF settings, the largest cross-validated joint RFF explained variance is $7.2\times10^{-4}$, which is far below the tolerance $\epsilon=0.01$. A few isolated probes are significant under the bootstrap test, but their joint out-of-fold explained variance is always very small. Based on the joint criterion (Section~\ref{subsec:diagnostics}), the held-out residuals show no clear remaining dependence on the retained modes within the tested RFF probe classes.

Finally, we evaluate the monotonicity diagnostics defined in Equations (\ref{eq:mono_delta}--\ref{eq:mono_summary}) on the final heat maps. As shown in Table~\ref{tab:monotonicity}, these decreases occur in a noticeable fraction of adjacent comparisons, especially for larger forcing values. Their sizes, however, are very small. The average violation is below $2\times10^{-5}$ for all forcings, and the maximum violation is at most $1.7\times10^{-3}$. Thus the violations are small local fluctuations, not a systematic failure of monotonicity. They are much smaller than the main recoverability differences in the heat maps and do not change the conclusions drawn from the maps.

\begin{table}[htbp]
    \centering
    \caption{Adjacent monotonicity-violation diagnostics for the final empirical recoverability maps. The diagnostics are computed over the $N_{\rm adj}=190$ adjacent admissible cutoff pairs $0\le \kcut\le k-2$, $k=1,\dots,20$.}
    \label{tab:monotonicity}
    \begin{ruledtabular}
    \begin{tabular}{@{}cccccc@{}}
        $F$ & $N_{\rm adj}$ & $N_{\rm viol}^{\rm mono}$ &
        $f_{\rm viol}^{\rm mono}$ & $V_{\max}^{\rm mono}$ &
        $V_{\rm avg}^{\rm mono}$ \\
        \colrule
        $8$ & $190$ & $39$ & $20.5\%$ & $1.2\times10^{-4}$ & $4.1\times10^{-6}$ \\
        $16$ & $190$ & $45$ & $23.7\%$ & $1.2\times10^{-4}$ & $6.2\times10^{-6}$ \\
        $32$ & $190$ & $53$ & $27.9\%$ & $1.2\times10^{-4}$ & $6.5\times10^{-6}$ \\
        $64$ & $190$ & $55$ & $28.9\%$ & $1.7\times10^{-3}$ & $1.6\times10^{-5}$ \\
    \end{tabular}
    
\end{ruledtabular}
\end{table}

As an additional baseline, Appendix~\ref{app:linear_baseline} compares the MLP estimator with an affine regressor for the pair $(k,\kcut)=(20,19)$. The affine recoverability is close to zero for all four forcing values. This shows that the reported high-mode recoverability does not come from linear correlation.

Together, these diagnostics support reading the reported maps as reliable estimates of the intrinsic recoverability.

\section{Conclusion}
\label{sec:discussion}

Applied to the Lorenz--96 system, the scale-resolved map $R(k\mid \kcut)$ reveals a finite high-mode band of partial conditional determination across $F=8,16,32,64$. The onset of substantial recoverability is organized around the direct quadratic triad-access scale $\lceil k/2\rceil$, while the overall amplitude decreases with forcing. Thus the map resolves where instantaneous coarse-to-fine deterministic information is present across target modes and cutoffs.

The RFF residual check supports this interpretation by testing whether fitted residuals still contain dependence on the retained modes. The diagnostic does not depend on the regression architecture, so it can also be used in other high-dimensional regression problems to test whether recoverable structure remains in a fitted residual. A natural next step is to apply the same map and diagnostic to other settings: \textcolor{purple}{wavelet truncation,} local-space or shell-energy observations in place of Fourier truncation, and larger systems.

\appendix
\section{Temporal decorrelation curves}
\label{app:decorrelation}

\begin{figure}[ht!]
    \centering
    \includegraphics[width=0.45\textwidth]{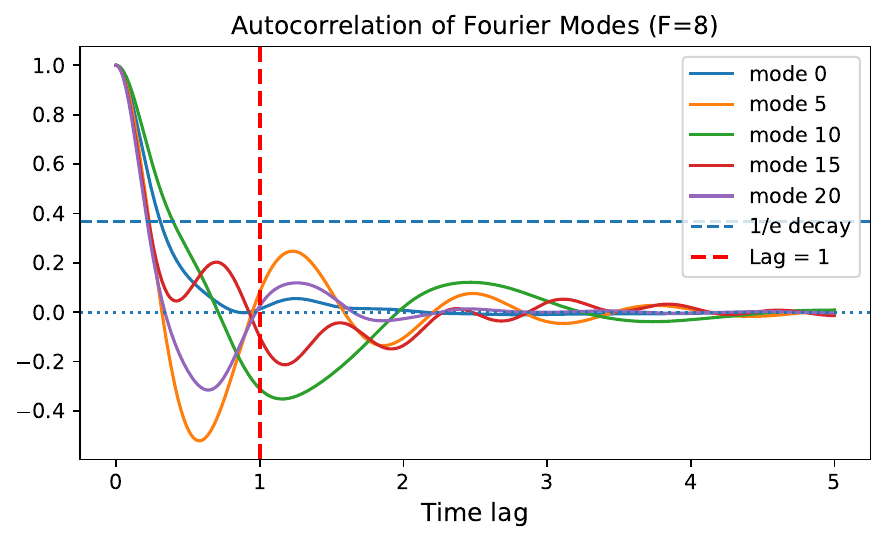}\hfill
    \includegraphics[width=0.45\textwidth]{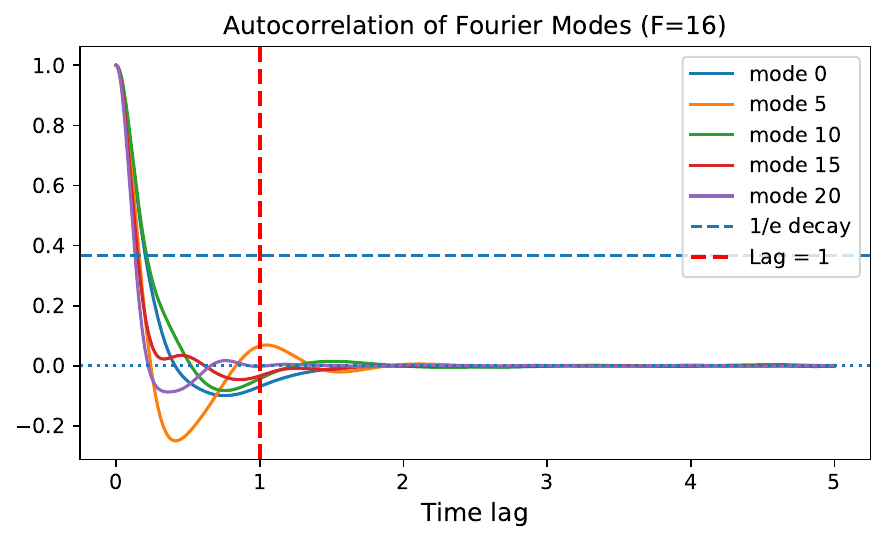}\\
    \includegraphics[width=0.45\textwidth]{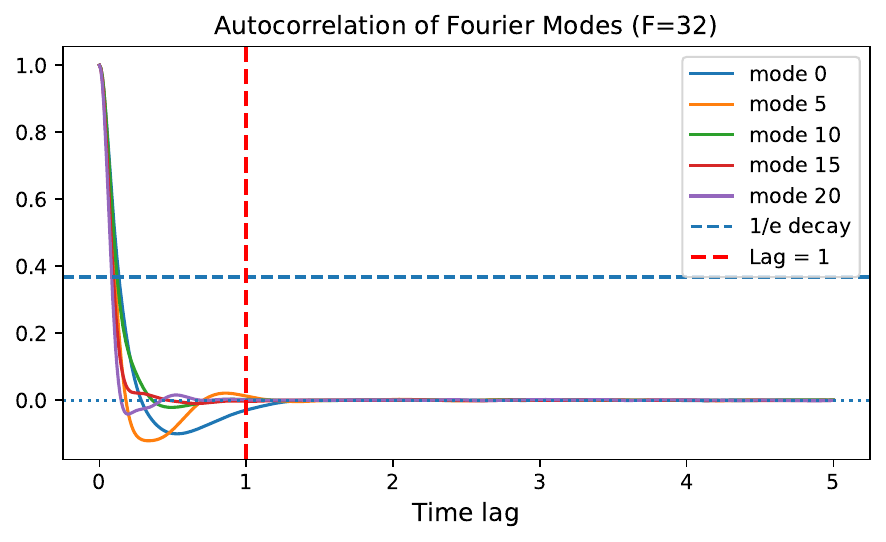}\hfill
    \includegraphics[width=0.45\textwidth]{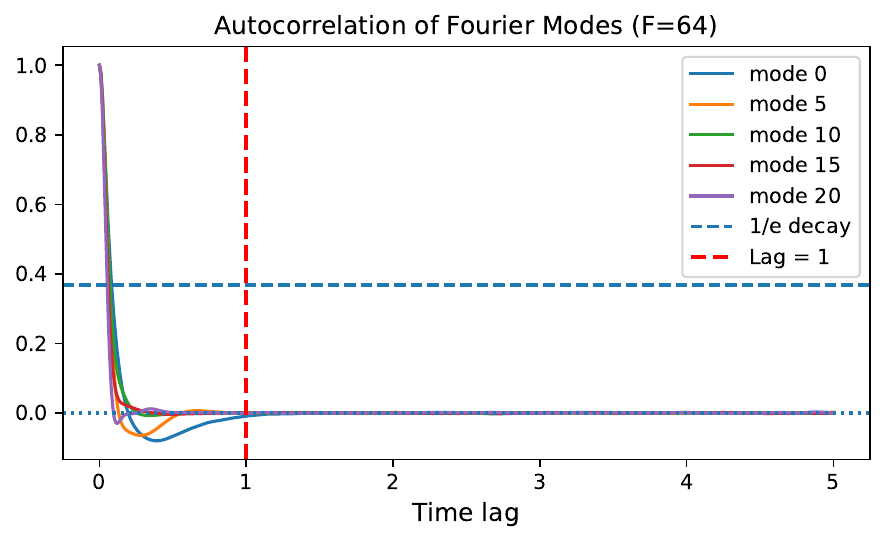}
    \caption{Autocorrelation curves for representative Fourier modes $k=0,5,10,15,20$ at $F=8,16,32,64$. The plotted quantity is $\operatorname{Re}\rho_k(\tau)$, defined in Equation~\eqref{eq:autocorr}. The horizontal dashed line marks the $1/e$ level. For all representative modes, $\operatorname{Re}\rho_k(\tau)$ falls below $1/e$ before $\tau=1$, and the decay is faster for stronger forcing.}
    \label{fig:decorrelation}
\end{figure}

Figure~\ref{fig:decorrelation} shows the full autocorrelation curves $\operatorname{Re}\rho_k(\tau)$ underlying the decorrelation times summarized in Table~\ref{tab:decorrelation}. The table reports only the scalar times $\tau_{1/e}$ and $\tau_{\rm int}$. The curves here show that the decay is rapid and largely monotone for all representative modes and forcings. Even at the weakest forcing, $F=8$, the autocorrelation falls below $e^{-1}$ before the sampling interval $\Delta t_{\rm samp}=1$. Thus the sampling interval is larger than the main short-time correlation scales, and neighboring retained snapshots are close to decorrelated.

\section{Random Fourier feature residual diagnostic}
\label{app:rff}

Here, we describe the details of the RFF residual-orthogonality diagnostic summarized in Section~\ref{subsec:diagnostics}. A fitted model equals the conditional mean exactly when its residual $e_k$ is orthogonal to every $\psi\in L^2(P_{Y_{\kcut}})$, as in Equation~\eqref{eq:orthogonality_condition_empirical}. To check this condition with finite data, we need to handle three issues. First, the condition is infinite-dimensional. It must hold for all square-integrable functions of the retained observation, including functions outside any fixed hand-chosen dictionary. We therefore test the residual with random features whose span becomes dense in $L^2(P_{Y_{\kcut}})$ as the number of probes increases. Second, an empirical correlation between one residual component and one probe can come from sampling noise. Since we test many probes and, for complex modes, multiple residual components at the same time, we calibrate the largest statistic over the full family. We use a multiplier bootstrap for this step, which keeps the empirical dependence among probes and components. Third, a statistically significant probe may still explain only a very small part of the residual variance. At the same time, real residual structure may be spread across many weak probes. For this reason, the main check is the joint out-of-fold explained variance over the full RFF span. The number of significant individual probes is reported only as a supporting quantity.

\paragraph{Probes.} Let $\{(Y_{\kcut}^{(i)},e_k^{(i)})\}_{i=1}^{n}$ be the held-out diagnostic set, where $e_k^{(i)}\in\R^{d_k}$ is the residual of the trained model and $n=|I_{\rm test}|$. We standardize each coordinate of the retained observation to have zero mean and unit variance, and write $\widetilde Y_{\kcut}^{(i)}$ for the standardized observation. This puts all retained coordinates on the same scale before we use Euclidean distances to build the probes. The residual is kept on its original scale. The probes are random Fourier features for a Gaussian kernel on the standardized observation space. We choose the median lengthscale $\gamma_{\rm med}$ by the median heuristic, namely the median pairwise Euclidean distance among the standardized observations. With this choice, the kernel bandwidth is close to a typical distance between retained states. A larger $\gamma$ gives smoother probes, while a smaller $\gamma$ probes finer changes in $Y_{\kcut}$. For a fixed lengthscale $\gamma$, we draw $w_j\sim N(0,\gamma^{-2}I)$ and $b_j\sim\mathrm{Unif}[0,2\pi]$, and define the probes $\psi_j$ as in \eqref{eq:rff_probe}. These are random features of the Gaussian kernel $\exp(-\|x-y\|_2^2/2\gamma^2)$~\cite{rahimi2007random}, whose spectral measure is $N(0,\gamma^{-2}I)$. As $D_m$ increases, their span becomes dense in $L^2(P_{Y_{\kcut}})$. Thus the diagnostic is not limited to a fixed finite dictionary or to a chosen polynomial degree.

\paragraph{Per-probe covariance statistics.} For residual component $c\in\{1,\dots,d_k\}$ and probe $j$, we define the empirical second moment of the probe, the residual--feature covariance, and the normalized covariance statistic by
\begin{equation}
\widehat v_j=\frac1n\sum_{i=1}^n\psi_j(\widetilde Y_{\kcut}^{(i)})^2,
\qquad
\widehat\mu_{c,j}=\frac1n\sum_{i=1}^n e_{k,c}^{(i)}\,\psi_j(\widetilde Y_{\kcut}^{(i)}),
\qquad
t_{c,j}=\frac{\sqrt n\,\widehat\mu_{c,j}}{\bigl[n^{-1}\sum_{i=1}^n (e_{k,c}^{(i)})^2\,\psi_j(\widetilde Y_{\kcut}^{(i)})^2\bigr]^{1/2}}.
\label{eq:app_perprobe}
\end{equation}
The statistic $t_{c,j}$ measures the size of the residual--probe covariance relative to its sampling scale. This normalization makes the statistics comparable across probes with different empirical sizes. For each probe, we record its largest signal over the residual components, $t_j^{\rm probe}=\max_c |t_{c,j}|$. Thus a probe is flagged if it has an unusually large covariance with at least one residual component.

\paragraph{Single-probe effect size.} The statistic $t_{c,j}$ is scaled by $\sqrt n$~\cite{chen2026diagnosing}, so it is useful for detecting nonzero covariance. However, it does not show how much residual variance is explained by the probe. We therefore also compute the single-probe effect size
\begin{equation}
e_j=\frac{\sum_{c=1}^{d_k}\widehat\mu_{c,j}^{\,2}}{\sigma_r^2\,\widehat v_j},
\qquad
\sigma_r^2=\frac1n\sum_{i=1}^n\|e_k^{(i)}\|_2^2.
\label{eq:app_effect}
\end{equation}
This quantity estimates the fraction of the total residual variance explained by probe $j$ alone. The factor $\widehat v_j$ is kept because a sampled probe does not have exactly unit norm under the empirical input distribution. The term $\sigma_r^2$ is the empirical total residual variance. Thus $t_j^{\rm probe}$ measures evidence against orthogonality for one probe, while $e_j$ measures the size of the corresponding single-probe effect.

\paragraph{Family-level threshold.} Since thousands of random probes are used, and since the probes and residual components are correlated, many large values of $t_{c,j}$ can appear by chance if each statistic is thresholded separately. We therefore set the threshold from the maximum statistic over all $(c,j)$ using a Gaussian multiplier bootstrap~\cite{chernozhukov2013gaussian}. Define the scores $g_{i,(c,j)}=e_{k,c}^{(i)}\psi_j(\widetilde Y_{\kcut}^{(i)})$ and $\bar g_{(c,j)}=n^{-1}\sum_i g_{i,(c,j)}$. For bootstrap replicate $b$, draw one multiplier $\xi_i^{(b)}\sim N(0,1)$ for each sample $i$, and use the same multiplier for all $(c,j)$ from that sample. Then compute
\begin{equation}
M_b=\max_{c,j}\left|\frac{1}{\sqrt n}\sum_{i=1}^n \xi_i^{(b)}\,\frac{g_{i,(c,j)}-\bar g_{(c,j)}}{s_{c,j}}\right|,
\qquad b=1,\dots,B,
\label{eq:app_bootstrap}
\end{equation}
where $s_{c,j}$ is the studentizing denominator in \eqref{eq:app_perprobe}. Using the same multiplier across all probes and components keeps their empirical dependence. Thus the bootstrap estimates the sampling distribution of the maximum statistic, not the distribution of one statistic alone. We take $c_\alpha$ as the $(1-\alpha)$ empirical quantile of $\{M_b\}_{b=1}^B$, and define the significant set as $S=\{j: t_j^{\rm probe}>c_\alpha\}$. This maximum-bootstrap threshold is less conservative than a Bonferroni correction when the probes are strongly correlated, while still controlling the family-level error for the residual--probe tests.

\paragraph{Joint criterion.} The primary diagnostic quantity is the fraction of residual variance explained by all RFF probes together. We first normalize each feature column by $\widehat v_j^{1/2}$, so that the ridge penalty treats all columns on the same scale. We then fit a $K$-fold cross-validated ridge regression from the normalized RFF features to the residual, and compute the out-of-fold explained variance $e_{\rm joint}^{\rm RFF}$ in \eqref{eq:rff_joint_effect}, clipped to $[0,1]$. The ridge penalty $\lambda_{\rm ridge}$ is chosen within cross-validation as the grid value that gives the largest $e_{\rm joint}^{\rm RFF}$. This quantity measures the effect size of the full probe span. It asks how much residual variance can still be predicted by nonlinear functions of the retained variables represented by the RFF features. The use of out-of-fold prediction is important. An in-sample explained variance can increase artificially as $D_m$ grows and the number of features approaches the sample size. The out-of-fold score avoids this issue because it does not reward overfitting. The ridge penalty grid includes large values, up to $10^6$, so that if the residual contains no remaining structure, cross-validation can choose a large penalty and drive $e_{\rm joint}^{\rm RFF}$ to zero instead of fitting noise.

\paragraph{Verdict and stability across settings.}
For one RFF diagnosis, we say that recoverable structure remains in the residual if
\begin{equation}
|S|>0 \quad\text{and}\quad e_{\rm joint}^{\rm RFF}>\epsilon.
\label{eq:app_verdict}
\end{equation}
This requires both at least one significant probe and a non-negligible joint explained variance. If either condition fails, the residual passes the check. Since one kernel bandwidth can only test part of the possible residual structure, we repeat the check over several settings. We use feature counts $D_m\in\{8000,16000,24000\}$ and three lengthscales $\gamma\in\{\gamma_{\rm med},\gamma_{\rm med}/2,\gamma_{\rm med}/4\}$. We then report the largest values of $|S|$ and $e_{\rm joint}^{\rm RFF}$ over all settings. We accept the residual as having no clear recoverable structure only when the conclusion is stable across all settings.

\paragraph{Mean-residual check.}
We also report the normalized mean residual as a simple bias check:
\begin{equation}
\widehat\mu_e(k,\kcut)=\frac{1}{|I_{\rm test}|}\sum_{i\in I_{\rm test}} e_k^{(i)}.
\label{eq:mean_residual}
\end{equation}
This quantity is computed after scaling the targets to $[0,1]$. It is the special case of Equation~\eqref{eq:orthogonality_condition_empirical} with $\psi\equiv 1$, so it is already covered by the RFF check. We still report it separately because a clearly nonzero mean residual is an easy-to-read sign of bias. For the representative pair $(\kcut,k)=(19,20)$, $\widehat\mu_e$ stays below $3\times 10^{-4}$ in magnitude for all four forcing values; see Table~\ref{tab:rff_residual_diagnostics}. This shows no clear detectable bias.

\section{Linear recoverability baseline for $(k, k_{\mathrm{cut}}) = (20, 19)$}
\label{app:linear_baseline}
To assess whether the recoverability measured in the main maps reflects genuine nonlinear conditional structure or merely linear correlation, we compute an affine least-squares baseline for the representative high-mode pair
$(k,\kcut)=(20,19)$ for all four forcings. The baseline defines an affine model
\[
g_{\mathrm{lin}}(Y_{19}) = Y_{19}W^\top + b
\]
fitted via ordinary least squares on the calibration block, with its performance
$\widehat{R}_{\mathrm{lin}}(20\mid19)$ subsequently evaluated on the held-out test block.

\begin{table}[h]
\begin{ruledtabular}
\centering
\caption{Linear recoverability baseline for $(k,\kcut)=(20,19)$. The affine least-squares model is fitted on the calibration block and
evaluated on the held-out test block. The MLP score is the final test
recoverability from the corresponding recoverability map. The Gap column
reports $\widehat{R}_{\mathrm{MLP}}-\widehat{R}_{\mathrm{lin}}$ on the
test set.}
\label{tab:linear_baseline}
\begin{tabular}{ccccc}
$F$ & $\widehat{R}_{\mathrm{lin}}$ (calib) & $\widehat{R}_{\mathrm{lin}}$ (test) & $\widehat{R}_{\mathrm{MLP}}$ (test) & Gap \\
\colrule
$8$  & $2.8\times10^{-5}$ & $-2.2\times10^{-5}$ & $0.8485$ & $0.8486$ \\
$16$ & $2.4\times10^{-5}$ & $-3.3\times10^{-5}$ & $0.6447$ & $0.6448$ \\
$32$ & $2.8\times10^{-5}$ & $-8.0\times10^{-6}$ & $0.4448$ & $0.4448$ \\
$64$ & $3.6\times10^{-5}$ & $-7.1\times10^{-5}$ & $0.2964$ & $0.2965$ \\
\end{tabular}
\end{ruledtabular}
\end{table}

As shown in Table~\ref{tab:linear_baseline}, the affine scores are indistinguishable from zero at all evaluated forcings. Both calibration and test values are of order $10^{-5}$--$10^{-4}$, with test values fluctuating slightly below zero. Consequently, $Y_{19}$ provides no appreciable linear conditional-mean information about $U_{20}$.

By contrast, the MLP recoverability for the same pair is large, ranging from $0.8485$ at $F=8$ to $0.2964$ at $F=64$. The gap between the affine and MLP scores shows that the recovered deterministic component is nonlinear. This is consistent with the quadratic triad coupling in the Lorenz--96 dynamics, where the retained modes can influence the target mode through products $\hat{x}_p\hat{x}_q$ satisfying $p+q\equiv k\pmod N$, rather than through a direct linear relation. The comparison supports the interpretation in Section~\ref{sec:results} that the recoverability map reflects nonlinear conditional dependence under the invariant measure, not linear correlation.

\bibliography{apssamp}

\end{document}